\RequirePackage{fix-cm}
\documentclass[smallextended]{svjour3}       
\smartqed  
\usepackage{graphicx}
\usepackage{latexsym,amssymb,amsmath,amsfonts}
\usepackage{algorithm,algorithmic}

\usepackage{latexsym,amssymb,amsmath,amsfonts}
\usepackage{theorem}
\usepackage{verbatim,array,multicol,palatino}
\usepackage{graphicx}
\usepackage{graphics}
\usepackage{fancyhdr}
\usepackage{algorithm,algorithmic}
\usepackage{url}

\def\thick#1{\hbox{\rlap{$#1$}\kern0.25pt\rlap{$#1$}\kern0.25pt$#1$}}

\def\lboxit#1{\vbox{\hrule\hbox{\vrule\kern6pt
      \vbox{\kern6pt#1\kern6pt}\kern6pt\vrule}\hrule}}
\def\thickboxit#1{\vbox{{\hrule height 1mm}\hbox{{\vrule width 1mm}\kern6pt
          \vbox{\kern6pt#1\kern6pt}\kern6pt{\vrule width 1mm}}
               {\hrule height 1mm}}}

\def\sM{{\mathcal M}}

\def\vectorfontone{\bf}
\def\vectorfonttwo{\boldsymbol}
\def\vd{{\vectorfontone d}}                      %
\def\vu{{\vectorfontone u}}                      
\def\vx{{\vectorfontone x}}                      
\def\vy{{\vectorfontone y}}                      
\def\vz{{\vectorfontone z}}                      %

\def\vone{{\vectorfontone 1}}
\def\vzero{{\vectorfontone 0}}

\def\vmu{{\vectorfonttwo \mu}}                   
\def\matrixfontone{\bf}
\def\matrixfonttwo{\boldsymbol}
\def\mI{{\matrixfontone I}}                      
\def\mX{{\matrixfontone X}}                      
\def\mZ{{\matrixfontone Z}}                      

\def\mSigma{{\matrixfonttwo \Sigma}}             %
\def\mPsi{{\matrixfonttwo \Psi}}                 %

\def\bP{{\mathbb P}}                             
\def\bI{{\mathbb I}}                             

\def\ds{\displaystyle}

\def\diag{\text{diag}}

\def\tr{\text{tr}}

\usepackage{amssymb,amsmath,amsfonts,palatino}
\usepackage{graphicx}

\usepackage{amssymb}

\usepackage[authoryear]{natbib}

\usepackage{latexsym,amssymb,amsmath,amsfonts}
\usepackage{theorem}
\usepackage{verbatim,array,multicol,palatino}
\usepackage{graphicx}
\usepackage{graphics}
\usepackage{fancyhdr}
\usepackage{algorithm,algorithmic}
\usepackage{url}

\usepackage{subfigure}
\usepackage{multirow}

\usepackage{geometry}
\def\ds{\displaystyle}

\def\bC{{\boldmath{C}}}
\def\bD{{\boldmath{D}}}
\def\bW{{\boldmath{W}}}
\def\bbeta{{\boldmath{\beta}}}
\def\ba{{\boldsymbol{\lambda}}}

\def\bu{{\boldmath{u}}}
\def\bc{{\boldmath{c}}}
\def\bx{{\boldmath{x}}}
\def\bX{{\boldmath{X}}}
\def\bZ{{\boldmath{Z}}}
\def\bzero{{\boldsymbol{0}}}
\def\bone{{\boldsymbol{1}}}

\def\bv{\boldmath{v}}

\def\ba{\boldsymbol{a}}
\def\bb{\boldsymbol{b}}
\def\bc{\boldsymbol{c}}
\def\bd{\boldsymbol{d}}

\def\br{\boldsymbol{r}}

\def\bu{\boldsymbol{u}}
\def\bv{\boldsymbol{v}}

\def\bx{\boldsymbol{x}}
\def\by{\boldsymbol{y}}
\def\bz{\boldsymbol{z}}

\def\bC{\boldsymbol{C}}
\def\bD{\boldsymbol{D}}

\def\bI{\boldsymbol{I}}

\def\bP{\boldsymbol{P}}
\def\bQ{\boldsymbol{Q}}

\def\bW{\boldsymbol{W}}
\def\bX{\boldsymbol{X}}
\def\bY{\boldsymbol{Y}}
\def\bZ{\boldsymbol{Z}}

\def\bbeta{\boldsymbol{\beta}}
\def\bgamma{\boldsymbol{\gamma}}

\def\bPsi{\boldsymbol{\Psi}}
\def\bGamma{\boldsymbol{\Gamma}}

\def\bzero{\boldsymbol{0}}
\def\bone{\boldsymbol{1}}

\def\tr{\mbox{tr}}

\def\bmu{\boldsymbol{\mu}}
\def\bSigma{\boldsymbol {\Sigma}}

\def\simind{\stackrel{{\tiny \mbox{ind.}}}{\sim}}

\def\btheta{\boldsymbol{\theta}}

\def\diag{\mbox{diag}}

\def\bOmega{\boldsymbol{\Omega}}
\def\bchi{\boldsymbol{\chi}}

\journalname{}

\begin{document}

\title{Mean Field Variational Bayesian Inference for \\
Support Vector Machine Classification}

\titlerunning{VB Inference for SVMs}        

\author{Jan Luts  \and
        John T. Ormerod}

\institute{Jan Luts \at
School of Mathematical Sciences, 
University of Technology, Sydney Broadway 2007, Australia \\
}

\date{Received: date / Accepted: date}

\maketitle

\begin{abstract}
\noindent A mean field variational Bayes approach to support vector 
machines (SVMs) using the latent variable representation on Polson \& Scott 
(2012) is presented. This representation allows circumvention of many of the
shortcomings associated with classical SVMs including automatic penalty 
parameter selection, 
the ability to handle dependent samples, missing data and variable selection. We 
demonstrate on simulated and real datasets that our approach is easily 
extendable to non-standard situations and outperforms the classical SVM approach 
whilst remaining computationally efficient.
\end{abstract}

\section{Introduction}

\noindent Support vector machines (SVMs) and its variants remain one of the most 
popular classification methods in machine learning and has been successfully 
utilized in many applications. Such applications include image classification, speech recognition, cancer diagnosis, natural language processing, forecasting, bio-informatics and as such these methods are likely to remain popular for 
many years to come. The strengths of SVMs derive from its formulation as an 
elegant convex optimization problem which can be efficiently solved, has few 
tuning parameters and whose solution only depends on a subset of the input
samples, called support vectors.

Despite such popularity standard SVMs suffer from several shortcomings. Section 
10.7 of \cite{HASTIE:2009} summarize these as: (i) natural handling data of 
mixed type, (ii) handling of missing values (iii) robustness to outliers in
input space (iv) insensitive to monotonic transformations of inputs (v) 
computational scalability to large sample sizes, (vi) inability to deal with
irrelevant inputs and (vii) intepretability. To this list we would add (viii)
the inability to deal with correlation within samples. In this paper we aim to
address (ii), (vi) and (viii).

This paper is not the first to consider these problems. Missingness has been 
considered by \cite{SMOLA:2005}, \cite{PELCKMANS:2005} and 
\cite{NEBOTTROYANO:2010}. Dealing with irrelevant inputs via variable/feature 
selection in SVMs has been considered by many authors including 
\cite{WESTON:2000}, \cite{TIPPING:2001}, \cite{GUYON:2002}, \cite{ZHU:2003}, 
\cite{GOLD:2005} and \cite{CHU:2006}. On the other hand, very few papers 
consider modification of SVMs to handle dependent or non-identically distributed
data. Notable exceptions include \cite{DUNDAR:2007}, \cite{LU:2011}, 
\cite{Pearce:2009} and \cite{LUTS:2012}. However, these problems are dealt with
in isolation and using different approaches, rather than in a unified manner
and it is difficult to see how these approaches could be adapted to multiple
complications, e.g., missingness and variable selection.

In the paper we follow the earlier work of \cite{BOSER:1992}, 
\cite{BISHOP:2000}, \cite{GAO:2005} and \cite{POLSON:2011} who propose various 
latent variable representations of the SVM loss function and reformulate the 
problem in a (pseudo-) Bayesian framework. This provides a unified approach 
which releases SVMs from many of the above problems including allowing efficient 
penalty parameter selection, correlation within samples, variable selection and 
missing data via well developed Bayesian methodology. Typically such Bayesian 
models are fit via Markov chain Monte Carlo (MCMC) methods. Unfortunately, MCMC 
methods can be notoriously slow when applied to large or complex models and can 
be rendered unsuitable in applications where speed is essential. These 
situations are precisely the same situations where SVMs are typically popular.

Our approach to this problem is to apply mean field variational Bayes (VB)
methods to the models we propose. The main advantage of this approach is a 
streamlined and computationally efficient framework for handling to many of the 
problems associated with the classical SVM approach. In tandem with these 
algorithms we also develop Gibbs sampling approaches to these methods to 
facilitate comparisons with an ``exact'' approach to these models.

In Section 2 we provide the framework for our approach. In Section 3 we consider
various extensions including automatic penalty parameter selection, group 
correlations, variable selection and missing predictors respectively. In Section 4
we show how our approach offers several computational advantages over the
classical SVM approach. In Section 5 we conclude. Appendices contain details of
our MCMC samplers.

\subsection*{Notation}

The notation $\bx\sim N(\bmu,\bSigma)$ means
that $\bx$ has a multivariate normal density with mean $\bmu$ and covariance 
$\bSigma$.  If $x$ has an inverse gamma distribution, denoted $x\sim 
\mbox{IG}(A,B)$, then it has density $p(x)=B^A\Gamma(A)^{-1}x^{-A-1}\exp(-B/x)$, 
$x,A,B>0$. If $x$ has an inverse Gaussian distribution, denoted \\
$x\sim \mbox{Inverse-Gaussian}(\mu,\lambda)$ with mean $\mu$ and variance 
$\mu^3/\lambda$, then it has density 
$$
p(x) = \sqrt{\frac{\lambda}{2\pi x^3}}
\exp\left\{ -\frac{\lambda(x - \mu)^2}{2x\mu^2}\right\}, 
\quad x,\mu,\lambda>0. 
$$

\noindent If $x$ has a generalized inverse Gaussian distribution, denoted 
$x\sim \mbox{GIG}(\gamma,\psi,\chi)$, then it has density 
$$
p(x) = \frac{(\psi/\chi)^{\gamma/2}}{2K_\gamma(\sqrt{\psi\chi})} 
x^{\gamma - 1}\exp\left\{ 
-\frac{1}{2}\left(\frac{\chi}{x}+\psi x\right)\right\}, 
\quad x,\psi,\chi>0, \ \gamma\in\mathbb{R},
$$

\noindent where $K_\gamma(\cdot)$ is a modified Bessel function of the second
kind. If $\bx$ is a vector of length $d$ then $\diag(\bx)$ is the $d\times d$ 
diagonal matrix whose diagonal elements are $\bx$. If $\bX$ is a $d\times d$ 
matrix then $\mbox{dg}(\bX)$ is the vector of length $d$ comprising of the 
diagonal elements of $\bX$. The $j$th column of a matrix $\mX$ is denoted $\mX_j$.

\section{Methodology} 

In this section we present a VB approach to a Bayesian SVM classification 
formulation for binary classification problems. After introducing Bayesian SVMs
and VB methodology we describe the latent variable SVM representation of 
\cite{POLSON:2011}  which gives rise to our basic VBSVM approach. 

\subsection{Bayesian support vector machines}
\label{sec:SVM}

Consider a training set $\{y_i,\bx_i\}_{i=1}^n$, where $\bx_i\in\mathbb{R}^p$ 
represents an input vector and $y_i\in\{-1,+1\}$ the corresponding class label. 
SVMs can be formulated in terms of finding a linear hyperplane that separates
the observations with $y_i = 1$ from those with $y_i = -1$ with the largest
minimal separating distance or margin. In general such a hyperplane does not 
exist and the problem needs to be reformulated as a trade-off between the size of the 
margin and infringements caused by points being on the wrong side of the 
hyperplane (for more details see for example \cite{VAPNIK:1998} or Chapter 12 of
\cite{HASTIE:2009}). This optimization problem amounts to finding 
$\bbeta\in\mathbb{R}^p$ which minimizes
\begin{equation}\label{eq:svmproblem}
\min_{\bbeta} {\mathcal{J}}(\bbeta) 
    = \left\{ \sum_{i=1}^n (1 - y_i \bx_i^T\bbeta)_+ \right\}
    + \alpha\|\bbeta\|^2,
\end{equation}

\noindent where $\alpha$ is a positive penalty parameter (the choice of which we 
will discuss later) and $x_+ = \max(0,x)$. Larger values of $\alpha$ serve to
shrink the fitted values of the $\bbeta$ coefficients. The above problem can be 
reformulated as a convex quadratic programming problem and can be solved using a
variety of efficient methods (for example Chapter 7 of \cite{Cristianini:2000}). 
This results in the classification rule $\mbox{sign}(\bx_i^T\bbeta)$ for input 
vector $\bx_i$. 

The terms $(1 - y_i\bx_i^T\bbeta)_+$ in (\ref{eq:svmproblem}) are referred to as
the hinge loss of the data and using a logarithmic scoring rule 
interpretation \citep{Bernardo:1979} can be interpreted as negative conditional
log-likelihoods. This has motivated Bayesian SVM formulations where
\begin{equation}\label{eq:bsvmproblem}
p\ell(y_i|\bbeta) = \exp\left\{
- (1 - y_i \bx_i^T\bbeta)_+ \right\}, \ 1\le i\le n, 
\quad \mbox{and} \quad 
\bbeta \sim N(\bzero,\tfrac{1}{2}\alpha^{-1}\bI_p),
\end{equation}

\noindent where $p\ell(y_i|\bbeta)$ is the pseudo-likelihood contribution of the
$i$th observation. Although (\ref{eq:bsvmproblem}) is not a true likelihood for
the remainder of the paper we will ignore this distinction and write 
$p\ell(y_i|\bbeta)$ as $p(y_i|\bbeta)$. Then 
$$
p(\by,\bbeta) 
= p(\bbeta)\prod_{i=1}^n p(y_i|\bbeta) 
\propto \exp\{ -{\mathcal{J}}(\bbeta) \},
$$

\noindent where $\by = [y_1,\ldots,y_n]^T$. Following \cite{Mallick:2005} we refer to formulations 
taking into account the normalizing constant of $p(\by,\bbeta)$ as complete
SVM (CSVM) formulations  whereas formulations ignoring the normalizing constant 
as Bayesian SVM (BSVM) formulations. We only consider the BSVM formulations 
here, i.e. (\ref{eq:bsvmproblem}).

Lastly, nonlinear classifiers can be constructed using kernelization methods
(see for example \cite{Zhang:2011}).

\subsection{Variational Bayesian inference}
\label{sec:VB1}

As discussed in the introduction the advantage of the Bayesian formulation is
that it allows us to extend SVM methodology to handle a variety of 
complications. Such Bayesian formulations are typically fit using MCMC
approaches \citep{Mallick:2005,POLSON:2011,Zhang:2011}. Unfortunately, MCMC 
approaches are often slow for the data mining applications where SVMs are 
typically used.

Mean field variational Bayes is a class of methods for approximate
Bayesian inference which are typically much faster than MCMC methods. Consider
the set of data ${\mathcal D}$ described by the joint likelihood 
$p({\mathcal D},\btheta)$ where $\btheta$ is a vector of parameters, latent
variables or missing values. Then it can be shown for a density of the form
$q(\btheta) = \prod_{i=1}^K q_i(\btheta_i)$ that the optimal $q_i^*(\btheta_i)$, 
which minimize the Kullback-Leibler distance between $q(\btheta)$ and 
$p(\btheta|{\mathcal D})$, satisfy
\begin{equation}\label{eq:vb}
q_i^*(\btheta_i) 
    \propto \exp\left[  
\mathbb{E}_{-q(\btheta_i)} \left\{ \log p({\mathcal D},\btheta) \right\}\right]
\end{equation}

\noindent where $\mathbb{E}_{-q(\btheta_i)}$ denotes expectations over
$\prod_{j\ne i} q_j^*(\btheta_j)$. If (\ref{eq:vb}) is calculated iteratively
over $i$ then the lower bound on the marginal log-likelihood 
\begin{equation}\label{eq:lb}
\log \underline{p}({\mathcal D};q) 
    = \mathbb{E}_{q(\btheta)} \left[ 
\log\left\{ \frac{p({\mathcal D},\btheta)}{q(\btheta)} \right\} \right]
\end{equation}

\noindent is guaranteed to increase monotonically. For more details and examples 
see \cite{Bishop:2006} or \cite{Ormerod:2010}.

\subsection{Variational Bayesian support vector machines} 
\label{sec:VB1andAHalf}

\cite{POLSON:2011} formulated an auxiliary variable representation of the 
problem analogous to (\ref{eq:svmproblem}), where the hinge loss is represented 
by a location-scale mixture of normal distributions. While \cite{Mallick:2005}
also consider an auxiliary representation of the SVM we use the 
representation of \cite{POLSON:2011}  because the calculation of 
(\ref{eq:vb}) and (\ref{eq:lb}) are analytically tractable.

Specifically, let 
$$
p(y_i,a_i|\bbeta) 
    = \frac{1}{\sqrt{2\pi a_i}} \exp\left[ -\frac{(1+a_i - y_i \bx_i^T\bbeta)^2}{2a_i} \right],
$$

\noindent where the $a_i>0$ values are auxiliary variables. Then the data 
augmentation approach of \cite{POLSON:2011} uses the fact that
$$
\exp[-2(1 - y_i \bx_i^T\bbeta)_+] = \int_0^{\infty} p(y_i,a_i|\bbeta) da_i.
$$

\noindent Hence, if $\bbeta\sim N(\bzero,\tfrac{1}{4}\alpha^{-1}\bI_p)$
then 
$$
\log p(\bbeta) 
+ \sum_{i=1}^n \left[ 
\log \int_0^{\infty} p(y_i,a_i|\bbeta) da_i \right] 
\propto -2{\mathcal{J}}(\bbeta). 
$$

\noindent Instead of performing inference on the parameter vector $\bbeta$ we 
treat $\ba = [a_1,\ldots,a_n]^T$ as random and perform inference on $\btheta=[\bbeta^T,\ba^T]^T$.
The main advantage of this representation is that the pseudo-conditional 
distribution $p(\by,\ba|\bbeta)=\prod_{i=1}^n p(y_i,a_i|\bbeta)$ is conjugate to
a multivariate normal distribution. 

Our proposed VB approach uses this representation of $p(\by,\ba|\bbeta)$ 
combined with the posterior density restriction 
$$
q(\bbeta,\ba)=q(\bbeta)\prod_{i=1}^{n} q(a_i).
$$ 

\noindent The resulting $q$-densities which minimize the Kullback-Leibler 
distance between $q(\ba,\bbeta)$ and the posterior densities are of the form
$$
\ds q^*(\bbeta) 
    \ds \sim N(\bmu_{q(\bbeta)},\bSigma_{q(\bbeta)}) 
\quad \mbox{and} \quad
q^*(a_i) 
    \simind \mbox{GIG}\left(\tfrac{1}{2},1,\chi_{q(a_i)} \right), 
\quad 1\le i\le n,
$$

\noindent where 
$$
\begin{array}{c}
\bSigma_{q(\bbeta)} 
    = \left( 
\bX^T\diag(\bmu_{q(\ba^{-1})})\bX + 4\alpha \bI_p
\right)^{-1}, \quad 
\bmu_{q(\bbeta)} 
    = \bSigma_{q(\bbeta)} \bX^T\bY(\bone_n + \bmu_{q(\ba^{-1})}), \\
\chi_{q(a_i)}
    = (1 - y_i\bx_i^T\bmu_{q(\bbeta)})^2 + \bx_i^T\bSigma_{q(\bbeta)}\bx_i
\quad \mbox{and} \quad \mu_{q(a_i^{-1})} = \chi_{q(a_i)}^{-1/2},
\ 1\le i\le n.
\end{array}
$$

In the above expressions $\bX$ denotes the $n$ by $p$ matrix such that the $i$th 
row of $\bX$ is $\bx_i$ and $\bY=\diag(\by)$. In order to reduce the length of 
later expressions we let $\bW = \diag(\bmu_{q(\ba^{-1})})$. The parameters 
$\bmu_{q(\bbeta)}$, $\bSigma_{q(\bbeta)}$ and $\chi_{q(a_i)}$ are determined by 
Algorithm 1. In Algorithm 1 the symbol $\odot$ denotes element-wise 
multiplication. 

Convergence of Algorithm 1 is monitored using the variational lower 
bound on the marginal likelihood, i.e., $\log\underline{p}(\by;q)$ given by 
$$
\begin{array}{rl}
\log\underline{p}(\by;q)
    & \ds = \tfrac{p}{2} 
          - n
          + n\log(2)
          - \tfrac{n}{2}\log(2\pi) 
          + \tfrac{p}{2}\log(4\alpha) 
          + \tfrac{1}{2}\log|\bSigma_{q(\bbeta)}|  
          - 2\alpha
\left[ \|\bmu_{q(\bbeta)}\|^2 + \mbox{tr}(\bSigma_{q(\bbeta)}) \right] \\
    & \ds \quad 
+ \by^T\bX\bmu_{q(\bbeta)} 
+ \tfrac{1}{4}\bone_n^T\log(\bchi_{q({\ba})}) 
+ \bone_n^T\log K_{1/2}(\sqrt{\bchi_{q({\ba})}}).
\end{array}
$$

\begin{algorithm}
\noindent \caption{\textit{Iterative scheme for obtaining the parameters in the optimal densities $q^*(\bbeta)$ and 
$q^*(\ba)$ for the variational Bayesian support vector machine with $\alpha$ fixed.}}
\label{alg1}
\begin{algorithmic}[1]
\REQUIRE $\bmu_{q({\ba^{-1}})} > \bzero$
\WHILE{the increase in $\log\underline{p}(\by;q)$ is significant}
    \STATE $\bW \leftarrow \diag(\bmu_{q(\ba^{-1})}) 
\quad ; \quad
\bSigma_{q(\bbeta)} \leftarrow \left( \bX^T\bW\bX + 4\alpha\bI_p \right)^{-1}
\quad ; \quad 
\bmu_{q(\bbeta)} \leftarrow \bSigma_{q(\bbeta)}\bX^T(\bI_n + \bW)\by$
    \STATE $\bchi_{q({\ba})} \leftarrow (\bone_n - \bY\bX\bmu_{q(\bbeta)})^2 
+ \mbox{dg}(\bX\bSigma_{q(\bbeta)}\bX^T)
\quad ; \quad\bmu_{q({\ba^{-1}})} \leftarrow \bchi_{q({\ba})}^{-1/2}$
\ENDWHILE
\end{algorithmic}
\end{algorithm}

\section{Extensions}

We now present a number of extensions addressing some of the shortcomings of
SVMs. These are presented in increasing complexity.
 
\subsection{Penalty parameter inference and random effect models} 
\label{sec:VB2}

Note that the positive penalization constant $\alpha$ in the previous section 
remains unspecified. By choosing an appropriate value for the constant $\alpha$
in (\ref{eq:svmproblem}), the objective function trades the loss term against 
the $\|\bbeta\|^2$ penalty term. This restricts the space of solutions, reduces 
the effect of overfitting and allows generalization to new, unseen data. Popular 
approaches for tuning the penalty parameter include cross-validation 
techniques and random sampling methods. However, these approaches to selecting 
$\alpha$ increase the overall computational overhead of these methods.

One approach to selecting the penalty parameter is by embedding a model into a 
mixed effect framework. Such an approach is commonly used to select the penalty 
parameter in penalized spline methods \citep{Wand:2003,Wand:2008}, the main 
by-product of which is enabling a natural embedding of semiparametric regression 
structures into SVM models \citep{Ruppert:2003,Zhao:2006}. 

We now show how the basic VBSVM model can be easily extended for selecting the 
penalty parameter automatically (without the need of a cross-validation 
strategy) and simultaneously how to handle group dependent data via a random 
intercept model. Let
\begin{equation}\label{model2}
\begin{array}{rl}
\ds p(\by,\ba|\bbeta,\bu) 
    & \ds = \exp\Big[
- n -\tfrac{n}{2}\log(2\pi) 
-\tfrac{1}{2}\bone_n^T\log(\ba)  
-\tfrac{1}{2}\bone_n^T(\ba + \ba^{-1}) \\
    & \ds \qquad +(\bone_n + \ba^{-1})^T\bY(\bX\bbeta + \bZ\bu) 
-\tfrac{1}{2}(\bX\bbeta + \bZ\bu) ^T\diag(\ba^{-1})(\bX\bbeta + \bZ\bu) 
\Big] \\
\quad 
\ds \mbox{and} \quad \bu|\bSigma
    & \ds \sim N(\bzero_m,\bSigma),
\end{array}
\end{equation}

\noindent where $\bX\in\mathbb{R}^{n\times p}$ and 
$\bZ\in\mathbb{R}^{n\times m}$ are fixed and random effect design matrices 
respectively and $\bSigma$ is the covariance of $\bu$. This class of models is 
extremely rich allowing random intercept, random slope, cross random effects, 
nested random effects, smoothing, generalized additive and semiparametric 
structures \citep{Zhao:2006}. We will consider the following two examples.

\medskip 
\noindent {\bf Example 1 [Penalty parameter selection]:} Consider the
data matrix $\bD\in\mathbb{R}^{n\times d}$ containing our observed predictors, 
i.e., $D_{ij}$ is the $i$th sample of the $j$th predictor. Suppose we wish to 
penalize the size of the coefficients associated with these predictors, but do 
not wish to penalize the size of the intercept coefficient. Then we would choose
$$
\bX = \bone_n, 
\quad \bZ = \bD
\quad \mbox{and} 
\quad \bSigma = \sigma_u^2\bI_m
$$

\noindent where $p = 1$, $m = d$ and $\sigma_u^2 = \frac{1}{4} \alpha^{-1}$.

\medskip 
\noindent {\bf Example 2 [Random intercept]:} Suppose we have the data 
$\{y_{i,j},\vd_{i,j}\}$, $1\le i\le m$, $1\le j\le n_i$ where $m$ is the number
of groups, $n_i$ is the number of observations in group $i$ and
$\vd_{i,j}\in\mathbb{R}^d$. Then we would define $n=\sum_{i=1}^m n_i$ and
choose
$$
\by = \left[ \begin{array}{c}
y_{1,1} \\
\vdots  \\
y_{1,n_1} \\
y_{2,1} \\
\vdots  \\
y_{m,n_m} \\
\end{array} \right], 
\quad
\bX = \left[ \begin{array}{cc}
1      & \vd_{1,1}^T \\
\vdots & \vdots  \\
1      & \vd_{1,n_1}^T \\
1      & \vd_{2,1}^T \\
\vdots & \vdots  \\
1      & \vd_{m,n_m}^T \\
\end{array} \right], 
\quad \bZ = \left[ \begin{array}{cccc}
\bone_{n_1}  & \bzero_{n_1} & \cdots & \bzero_{n_1} \\
\bzero_{n_2} & \bone_{n_2}  & \cdots & \bzero_{n_2} \\
\vdots       & \vdots       & \ddots & \vdots \\
\bzero_{n_m} & \bzero_{n_m} & \cdots & \bone_{n_m} \\
\end{array} \right]
\quad \mbox{and} 
\quad \bSigma = \sigma_u^2\bI_m
$$

\noindent where $p = d + 1$ and $\sigma_u^2$ is the random intercept variance.

For both these examples we use the priors
$$
\ds \bbeta \sim N(\vzero_p,\sigma_\beta^2\bI_p)
\quad \mbox{and} \quad 
\sigma_u^2\sim \mbox{IG}(A_u,B_u)
$$

\noindent where $\sigma_\beta^2$, $A_u$ and $B_u$ are fixed prior 
hyperparameters. For our numerical experiments we use the values 
$\sigma_\beta^2 = 10^8$ and $A_u = B_u = 0.01$ as defaults to impose non-informativity.
Placing a prior on $\sigma_u^2=\alpha^{-1}/4$ 
allows us to perform inference on the penalty parameter.

To apply the VB method we choose the product restriction for approximating the 
posterior density of the form
$$
q(\bbeta,\bu,\sigma_u^2,\ba) 
= q(\bbeta,\bu) q(\sigma_u^2) \prod_{i=1}^{n} q(a_i). 
$$

\noindent Using (\ref{eq:vb}) and this product restriction the optimal $q$-densities are of the form
$$
\begin{array}{c}
\ds q^*(\bbeta,\bu) 
    \ds \sim N(\bmu_{q(\bbeta,\bu)},\bSigma_{q(\bbeta,\bu)}), \quad
q^*(\sigma_{u}^{2}) 
    \ds \sim \mbox{IG}\left( A_u + \tfrac{m}{2}, B_{q(\sigma_{u}^{2})} \right) 
\\  
\mbox{and} \quad
q^*(a_i) 
    \simind \mbox{GIG}\left(\tfrac{1}{2},1,\chi_{q(a_i)} \right), 
\quad 1\le i\le n,
\end{array}
$$

\noindent where the parameters are determined by Algorithm 2. In Algorithm 2
the matrix $\bC$ is $[\bX,\bZ]$ and in the main loop the lower bound 
on the marginal log-likelihood simplifies to
$$
\begin{array}{rl}
\ds \log\underline{p}(\by;q) 
    & \ds = \tfrac{p+m}{2} 
          - n
          + n\log(2)
          - \tfrac{n}{2}\log(2\pi) 
          - \tfrac{p}{2}\log(\sigma_\beta^2) 
          + \tfrac{1}{2}\log|\bSigma_{q(\bbeta,\bu)}|
          - \tfrac{1}{2\sigma_\beta^{2}}\left[
\|\bmu_{q(\bbeta)} \|^2 + \mbox{tr}(\bSigma_{q(\bbeta)}) \right] \\
    & \ds \quad + A_u\log(B_u) 
                - \log\Gamma(A_u)
                - \left( A_u + \tfrac{m}{2}\right)\log(B_{q(\sigma_u^{2})}) 
                + \log\Gamma\left( A_u + \tfrac{m}{2} \right) \\
    & \ds \quad 
+ \by^T\bC\bmu_{q(\bbeta,\bu)}
+ \tfrac{1}{4}\bone_n^T\log(\bchi_{q({\ba})}) 
+ \bone_n^T\log K_{1/2}(\sqrt{\bchi_{q({\ba})}}),
\end{array}
$$

\noindent where $\Gamma(\cdot)$ is the gamma function. Classification of a new input vector $\bc_i$ is performed based on the
value $\text{sign}(\bc_i^T\bmu^*_{q(\bbeta,\bu)})$.

\begin{algorithm}
\noindent \caption{\textit{Iterative scheme for obtaining the parameters in the
optimal densities $q^*(\bbeta,\bu)$, $q^*(\sigma_{u}^{2})$ and 
$q^*(\ba)$ for the variational Bayesian support 
vector machine with penalty parameter or random intercept inference.}}
\label{alg2}
\begin{algorithmic}[1]
\REQUIRE $\bmu_{q({\ba^{-1}})} > \bzero_n, \mu_{q({\sigma_{u}^{-2}})}>0$
\WHILE{the increase in $\log\underline{p}(\by;q)$ is significant}
    \STATE $\bW \leftarrow \diag(\bmu_{q({\ba^{-1}})})
\quad ; \quad
\bSigma_{q(\bbeta,\bu)} \leftarrow \left[ \bC^T\bW\bC + 
\mbox{blockdiag}(\sigma_\beta^{-2}\bI_p,\mu_{q({\sigma_{u}^{-2}})}\bI_m) \right]^{-1}$ 
    \STATE $\bmu_{q(\bbeta,\bu)} \leftarrow 
\bSigma_{q(\bbeta,\bu)}\bC^T(\bI_n + \bW)\by
\quad ; \quad
\bchi_{q({\ba})} \leftarrow (\bone_n - \bY\bC\bmu_{q(\bbeta,\bu)})^2 
+ \mbox{dg}(\bC\bSigma_{q(\bbeta,\bu)}\bC^T)
\quad ; \quad
\bmu_{q({\ba^{-1}})} \leftarrow \bchi_{q({\ba})}^{-1/2}$
    \STATE $B_{q(\sigma_u^2)} \leftarrow 
B_u + \tfrac{1}{2}\left[\|\bmu_{q(\bu)}\|^2 + \text{tr}(\bSigma_{q(\bu)})\right] 
\quad ; \quad 
\mu_{q({\sigma_u^{-2}})} \leftarrow (A_u+m/2)/B_{q(\sigma_u^2)}$
\ENDWHILE
\end{algorithmic}
\end{algorithm}

\subsection{Variable selection}
\label{sec:VB3}

The classical 2-norm SVM and the VBSVM approaches we have described so far 
include no mechanism to induce sparsity for the fitted coefficients. Inducing
sparsity is important because it allows us to remove potentially irrelevant or 
redundant variables. In this section we present a VBSVM which overcomes this
via the incorporation of a sparse prior. While there exist numerous options for
the choice of the sparse prior, we will make use of the Laplace-zero density 
\citep{Wand:2011}. 

Consider again the model (\ref{model2}). Suppose that we wish the fitted 
$\bbeta$ be included in the model (which may include terms such as the 
intercept) whereas we wish to induce sparsity on the vector of fitted $\bu$. 
Instead of using $\bu|\sigma_u^2 \sim N(\bzero_m,\sigma_u^2\bI_m)$ consider the 
hierarchical prior for $u_k$ given by
$$
\begin{array}{c}
u_k|\gamma_k,\sigma_u
\simind \gamma_k \mbox{Laplace}(0,\sigma_u)  
+ (1 - \gamma_k)\delta_0, 
\quad
\gamma_k|\rho \simind \mbox{Bernoulli}(\rho),
\quad 1\le k\le m, \\
\end{array}
$$

\noindent where $\delta_0$ is the degenerate distribution with point mass at 
$0$. This representation is unnatural to work with using VB methodology and so
we use a more natural representation.

First, we note that the Laplace distribution can be represented using a 
normal-scale mixture \citep{Andrews:1974}. This representation uses the
fact that  
$$
\mbox{if} \quad v_k|b_k,\sigma_u^2 \simind N(0,\sigma_u^2/b_k) 
\quad \mbox{and} \quad
b_k \simind \mbox{IG}(1,1/2) \quad \mbox{then} \quad
v_k|\sigma_u\stackrel{\mbox{\tiny ind.}}{\sim} 
\mbox{Laplace}(0,\sigma_u).
$$

\noindent Hence, instead of (\ref{model2}) we let $u_k = \gamma_k v_k$ and use
$$
\begin{array}{c}
\begin{array}{rl}
\ds p(\by,\ba|\bbeta,\bv,\bgamma) 
    & \ds = \exp\Big[
-n
-\tfrac{n}{2}\log(2\pi) 
-\tfrac{1}{2}\bone_n^T\log(\ba)  
-\tfrac{1}{2}\bone_n^T(\ba + \ba^{-1}) \\
    & \ds \qquad + 
(\bone_n + \ba^{-1})^T\bY(\bX\bbeta + \bZ\bGamma\bv) 
- \tfrac{1}{2}
(\bX\bbeta + \bZ\bGamma\bv)^T\diag(\ba^{-1})(\bX\bbeta + \bZ\bGamma\bv) 
\Big],
\end{array} \\
\ds \bv|\bb,\sigma_u^2 \sim N(\bzero,\sigma_u^2\diag(\bb^{-1})), \quad
\quad b_k \simind \mbox{IG}(1,1/2)
\quad \mbox{and}
\quad \gamma_k \simind \text{Bernoulli}(\rho), \quad 1 \leq k \leq m,
\end{array}
$$

\noindent where $\bb = [b_1,\ldots,b_m]^T$, $\bv=[v_1,\ldots,v_m]$, $\bgamma =[\gamma_1,\ldots,\gamma_m]$ and
$\bGamma = \diag(\bgamma)$, and use the priors
$$
\bbeta \sim N(\bzero_p,\sigma_\beta^2\bI_p) 
\quad \mbox{and} \quad
\quad \sigma_u^2\sim \mbox{IG}(A_u,B_u).
$$
\noindent The set of auxiliary variables $\gamma_k$ and $b_k$ has been 
introduced such that $u_k|\gamma_k,\sigma_u$ has the desired Laplace-zero 
prior distribution. The hyperparameter $\rho$ is chosen in function of the desired level of sparsity.

Next, a factorization is specified for the approximation to the posterior 
density function by
$$
q(\bbeta,\bv,\sigma_u^2,\ba,\bb,\bgamma) = 
q(\bbeta,\bv) q(\sigma_u^2) \left[ \prod_{i=1}^n q(a_i) \right]
\left[ \prod_{k=1}^m q(b_k)q(\gamma_k) \right].
$$
\noindent Using (\ref{eq:vb}) and this product restriction the optimal 
$q$-densities are of the form
$$
\begin{array}{c}
q^*(\bbeta,\bv) \sim N(\bmu_{q(\bbeta,\bv)},\bSigma_{q(\bbeta,\bv)}),
\quad
q^*(\sigma_u^{2}) \sim \mbox{IG}\left(A_u+\frac{m}{2},B_{q(\sigma_u^2)}\right), \\ 
q^*(b_k) \simind \mbox{Inverse-Gaussian}\left(\mu_{q(b_k)},1\right), \quad
q^*(\gamma_k) \simind \mbox{Bernoulli}\left( \mu_{q(\gamma_k)}\right), \ 1\le k\le m, \\
\mbox{and} \quad q^*(a_i) \simind \mbox{GIG}(\frac{1}{2},1,\chi_{q(a_i)}), \ 1 \le i\le n,
\end{array}
$$

\noindent where the parameters are determined by Algorithm \ref{alg3}. In 
Algorithm \ref{alg3} we use the function $\mbox{expit}(x) = 1/(1+\exp(-x))$, let 
$\bZ_k$ denote the $k$th column of $\bZ$ and let $\bZ_{-k}$ be the $\bZ$ matrix 
with the $k$th column removed.
The lower bound in the main loop in Algorithm \ref{alg3} takes the 
simplified form
$$
\begin{array}{rl}
\ds \log\underline{p}(\by;q) 
    & \ds =
(n - m)\log(2)
- n 
+ \tfrac{p+m}{2} 
- \tfrac{n - m}{2}\log(2\pi) 
\\
    & \ds \quad 
+ \by^T\bC\diag(\bmu_{q(\widetilde{\bgamma})})\vmu_{q(\bbeta,\bv)}
+ \tfrac{1}{4}\bone_n^T\log(\bchi_{q(\ba)})
+ \bone_n^T\log K_{1/2}(\sqrt{\bchi_{q(\ba)}})
\\
    & \ds \quad 
+ \tfrac{1}{2}\log|\bSigma_{q(\bbeta,\bv)}|
- \tfrac{p}{2}\log(\sigma_\beta^2) 
- \tfrac{1}{2\sigma_\beta^2}
\left[ \|\bmu_{q(\bbeta)}\|^2 + \tr(\bSigma_{q(\bbeta)})\right]
- \tfrac{1}{2}\bone_m^T\bmu_{q(\bb)}^{-1}
\\
    & \ds \quad 
+ A_u\log(B_u) 
- \log\Gamma(A_u) 
- \left( A_u + \tfrac{m}{2} \right)\log(B_{q(\sigma_u^2)}) 
+ \log\Gamma\left( A + \tfrac{m}{2} \right)
\\
    & \ds \quad 
- \bmu_{q(\bgamma)}^T\log\left(\frac{\bmu_{q(\bgamma)}}{\rho\bone_m}\right) 
- (\bone_m - \bmu_{q(\bgamma)})^T\log\left(\frac{\bone_m - \bmu_{q(\bgamma)}}{(1- \rho)\bone_m}\right),
\end{array}
$$

\noindent where $\widetilde{\bgamma} = [\bone_p^T,\bgamma^T]^T$ and 
$\bC = [\bX,\bZ]$.
The converged solutions $\bmu^*_{q(\bbeta,\bv)}$ and
$\bmu^*_{q(\widetilde{\bgamma})}$ allow the establishment of the classification rule 
$\text{sign}(\bc_i^T(\bmu^*_{q(\bbeta,\bv)}\odot\bmu^*_{q(\widetilde{\bgamma})}))$ 
for classifying input vector $\bc_i$. The vector 
$\bmu^*_{q(\widetilde{\bgamma})} = [\vone_p^T,\bmu^{*^T}_{q(\bgamma)}]^T$ provides 
probability measures to decide which of the original input variables to select. 

\begin{algorithm}
\noindent \caption{\textit{Iterative scheme for obtaining the parameters in the 
optimal densities for the variational Bayesian support vector machine with 
Laplace-zero prior.}}
\label{alg3}
\begin{algorithmic}[1]
\REQUIRE $\bmu_{q({\ba^{-1}})},\mu_{q({\sigma_{u}^{-2}})},\bmu_{q(\bb)}, 
\bmu_{q(\widetilde{\bgamma})}, \bOmega_{q(\widetilde{\bgamma})}$
\WHILE{the increase in $\log\underline{p}(\by;q)$ is significant}
    \STATE $\bW \leftarrow \diag(\bmu_{q(\ba^{-1})})
\quad ; \quad \bSigma_{q(\bbeta,\bv)} \leftarrow \left[ 
(\bC^T\bW\bC)\odot\bOmega_{q(\widetilde{\bgamma})} 
    + \mbox{blockdiag}
(\sigma_{\beta}^{-2}\bI_p,\mu_{q(\sigma_u^{-2})}\diag(\bmu_{q(\bb)})) 
\right]^{-1}$
    \STATE $\bmu_{q(\bbeta,\bv)} \leftarrow \bSigma_{q(\bbeta,\bv)} \diag(\bmu_{q(\widetilde{\bgamma})})\bC^T(\bI_n+\bW)\by 
\quad ; \quad 
\bOmega_{q(\bbeta,\bv)} \leftarrow 
\bSigma_{q(\bbeta,\bv)} + \bmu_{q(\bbeta,\bv)}\bmu_{q(\bbeta,\bv)}^T$
    \FOR{$k=1,\ldots,m$}
        \STATE $\mu_{q(b_k)} \leftarrow 
\left[ \mu_{q(\sigma_{u}^{-2})} \bOmega_{q(v_k,v_k)} \right]^{-1/2}$
        \STATE $\begin{array}{rl} 
\ds \eta_{q(\gamma_k)} 
    & \leftarrow 
\mbox{logit}(\rho) 
- \tfrac{1}{2}\bZ_k^T\bW\bZ_k \bOmega_{q(v_k,v_k)} 
+ \bZ_k^T\by\mu_{q(v_k)}  \\
    & \ds \qquad 
+ \bZ_k^T\bW(\by\mu_{q(v_k)} 
- \bX\bOmega_{q(\bbeta,v_k)} 
- \bZ_{-k}\diag(\bmu_{q(\bgamma_{-k})})\bOmega_{q(\bv_{-k},v_k)})
\end{array}$
        \STATE $\mu_{q(\gamma_k)} \leftarrow \mbox{expit}(\eta_{q(\gamma_k)})$
    \ENDFOR
    \STATE $\bmu_{q(\widetilde{\bgamma})} \leftarrow [\bone_p^T,\bmu^T_{q(\bgamma)}]^T
\quad ; \quad 
\bSigma_{q(\widetilde{\bgamma})} \leftarrow
\diag(\bmu_{q(\widetilde{\bgamma})}\odot(\bone_{(p+m)}-\bmu_{q(\widetilde{\bgamma})}))
\quad ; \quad
\bOmega_{q(\widetilde{\bgamma})} \leftarrow \bSigma_{q(\widetilde{\bgamma})} + \bmu_{q(\widetilde{\bgamma})}\bmu_{q(\widetilde{\bgamma})}^T$
     \STATE $\bchi_{q({\ba})} \leftarrow
\bone_n - 2\bY\bC\diag(\bmu_{q(\widetilde{\bgamma})})\bmu_{q(\bbeta,\bv)}
+ \mbox{dg}
\{ \bC(\bOmega_{q(\widetilde{\bgamma})} \odot \bOmega_{q(\bbeta,\bv)})\bC^T\} 
\quad ; \quad
\bmu_{q(\ba^{-1})} \leftarrow \bchi_{q({\ba})}^{-1/2}$
    \STATE $B_{q(\sigma_u^2)} 
\leftarrow B_u + \tfrac{1}{2} 
\mbox{tr}\left[ \diag(\bmu_{q(\bb)})
(\bSigma_{q(\bv)} + \bmu_{q(\bv)}\bmu_{q(\bv)}^T) \right] 
\quad ; \quad 
\mu_{q({\sigma_u^{-2}})} \leftarrow (A_u+m/2)/B_{q(\sigma_u^2)}$
\ENDWHILE
\end{algorithmic}
\end{algorithm}
 
\subsection{Missing predictor values}
\label{sec:VB4}

The last extension we present in this paper provides the methodology to deal 
with the situation where there exist missing values in the training data vectors 
$\vd_i$. The classical SVM formulation in Section \ref{sec:SVM} requires 
training input vectors which are completely observed. Similarly, the VB 
approaches in Section \ref{sec:VB1andAHalf}-\ref{sec:VB3} don't allow any missing 
values. This section outlines a missing data extension for the penalty  
parameter inference methodology from Section \ref{sec:VB2}.

%

For missing data situations we consider data triple $\{y_i,\vd_i,\br_i \}$
for the $i$th sample with $1\le i\le n$. Here $y_i\in\{-1,+1\}$ is the $i$th 
response, $\vd_i\in\mathbb{R}^d$ is the $i$th vector of predictors 
and $\br_i$ is an indicator vector where $r_{ij}=1$ if the $j$th predictor of the $i$th input vector is observed and $0$ otherwise.

We will assume that the likelihood for $\{y_i,\vd_i,\br_i\}$ factorizes as
$$
p(y_i,\vd_i,\br_i) = p(y_i|\vd_i)p(\vd_i) p(\br_i|\vd_i).
$$

\noindent In terms of missing data jargon this is called a selection model. We
will refer to $p(y_i|\vd_i)$, $p(\vd_i)$ and $p(\br_i|\vd_i)$ as the regression,
imputation and missing data mechanism components of the model respectively. If 
$\br_i$ and $\vd_i$ are independent so that $p(\br_i|\vd_i) = p(\br_i)$ then we 
say that the data are missing completely at random (MCAR). Let 
$\mathcal{J} = \{ j \colon r_{ij} = 1 \ \mbox{for all} \ 1\le i\le n \}$.
If $p(\br_i|\vd_i) = p(\br_i|\vd_{i,\mathcal{J}})$, i.e., $\br_i$ depends on 
completely observed predictors then the data is missing at random (MAR). 
Finally, in the general case the missingness depends on the data and we say that
the data is missing not at random (MNAR). In the MCAR and MAR cases inferences
for parameters in the regression and imputation components can be performed
independently of inferences for parameters in the missing data mechanism and
we say that the missing data mechanism is ignorable. For simplicity we will
assume the data are MCAR. The MAR and MNAR cases can be adapted from \cite{Faes:2011}.

Consider the regression component of the model 
$$
\begin{array}{c}
\begin{array}{rl}
\ds p(\by,\ba|\bbeta,\bu,\bD) 
    & \ds = \exp\Big[
-n
-\tfrac{n}{2}\log(2\pi) 
-\tfrac{1}{2}\bone_n^T\log(\ba)  
-\tfrac{1}{2}\bone_n^T(\ba + \ba^{-1}) \\
    & \ds \qquad 
+ (\bone_n + \ba^{-1})^T\bY(\vone_n\beta + \bD\bu) 
-\tfrac{1}{2}(\vone_n\beta + \bD\bu)^T\diag(\ba^{-1})(\vone_n\beta + \bD\bu) 
\Big], \\
\end{array} \\
\begin{array}{c}  
\beta \sim N(0,\sigma_\beta^2),
\quad \bu|\sigma_u^2 \sim N(\bzero_d,\sigma_u^2\bI_d),
\quad \sigma_u^2 \sim \mbox{IG}(A_u,B_u),
\end{array}
\end{array}
$$

\noindent where $\vd_i$ are stored in the rows of $\bD\in\mathbb{R}^{n\times d}$ and we model the imputation model via 
$$
\begin{array}{c} 
\vd_i|\bmu,\bSigma \simind N(\bmu,\bSigma),\quad 1 \leq i \leq n,
\quad \bmu\sim N(\bzero_d,\sigma_\mu^2\bI_d)
\quad \mbox{and} \quad \bSigma\sim\mbox{IW}(\bPsi,\nu),
\end{array}
$$

\noindent where $\mbox{IW}(\bPsi,\nu)$ denotes the inverse Wishart
distribution with scale matrix $\bPsi$ and degrees of freedom $\nu$. In our 
examples we use $\sigma_\mu^2 = 10^8$, $\bPsi = 0.01 \, \bI_d$ and $\nu= 3$.


Let $\mathcal{M}=\{i\colon\br_i^T\bone_d\ne d\}$,
$\mathcal{M}_i=\{j\colon r_{ij} = 0\}$ and $\bD_{\mbox{\tiny mis}}$ and $\bD_{\mbox{\tiny obs}}$ denote the 
components of $\bD$ that are missing and observed respectively. Then we approximate the posterior density 
using the factorization
$$
q(\beta,\bu,\sigma_u^2,\ba,\bD_{\mbox{\tiny mis}},\bmu,\bSigma) 
= q(\beta,\bu) q(\sigma_u^2) \left[ \prod_{i=1}^n q(a_i) \right]
\left[ \prod_{i\in\mathcal{M}} q(\vd_{i,\mathcal{M}_i}) \right]
q(\bmu) q(\bSigma).
$$

\noindent Using (\ref{eq:vb}) and this product restriction the optimal 
$q$-densities are of the form
$$
\begin{array}{c}
\ds q^*(\beta,\bu) 
    \sim N(\bmu_{q(\beta,\bu)},\bSigma_{q(\beta,\bu)}), \quad
q^*(a_i) 
    \simind \mbox{GIG}(\tfrac{1}{2},1,\chi_{q(a_i)}), \quad 
q^*(\sigma_u^2) 
    \sim \mbox{IG}(A_u + \tfrac{d}{2},B_{q(\sigma_u^2)}), \\
q^*(\bmu) 
    \sim N(\bmu_{q(\bmu)},\bSigma_{q(\bmu)}), \quad
q^*(\bSigma) 
    \sim \mbox{IW}(\bPsi_{q(\bSigma)},\nu + n) \quad \mbox{and} \quad
q^*(\vd_{i,\mathcal{M}_i}) 
    \simind 
N(\bmu_{q(\vd_{i,\mathcal{M}_i})},\bSigma_{q(\vd_{i,\mathcal{M}_i})}).
\end{array}
$$

\noindent Finally, Algorithm \ref{alg5} combines these component-wise solutions
in an iterative scheme to obtain the simultaneous solution for VBSVM classification with missing values. To reduce space we have used the following notation:
let $\bP_i$ be the $d$ by $|\mathcal{M}_i|$ matrix consisting of the columns of
$\bI_d$ with indices $\mathcal{M}_i$ and let $\bQ_i$ be the $d$ by $(d-|\mathcal{M}_i|)$ matrix consisting of the remaining columns of
$\bI_d$. If $|\mathcal{M}_i| = 0$ then $\bP_i = \bzero_d$ and if $|\mathcal{M}_i| = d$ then $\bQ_i = \bzero_d$. Let $\widetilde{\bC}$ be the
$n\times (1+d)$ matrix such that the $i$th row of $\widetilde{\bC}=\vmu_{q(\bC)}$ is given by
$$
\widetilde{\bc}_i 
= \bmu_{q(\bc_i)} 
= [1,(\bQ_i\bQ_i^T\vd_i + \bP_i\bmu_{q(\vd_{i,\mathcal{M}_i})})^T]^T.
$$

\noindent The lower bound in Algorithm \ref{alg5} takes the form
$$
\begin{array}{rl}
\log\underline{p}(\by,\bD_{\mbox{\tiny obs}};q) 
    & \ds =
n \log(2)
- n 
- \tfrac{n}{2}\log(2\pi)
+ \by^T\widetilde{\bC}\bmu_{q(\beta,\bu)}
+ \tfrac{1}{4}\bone_n^T\log(\bchi_{q(\ba)})
+ \bone_n^T\log K_{1/2}(\sqrt{\bchi_{q(\ba)}})\\
    & \ds \quad 
+ A_u\log(B_u) 
- \log\Gamma(A_u) 
- \left( A_u + \tfrac{d}{2} \right)\log(B_{q(\sigma_u^2)}) 
+ \log\Gamma\left( A_u + \tfrac{d}{2} \right) \\
    & \ds \quad 
- \tfrac{1}{2}\log(\sigma_\beta^2) 
- \tfrac{1}{2\sigma_\beta^2}
\left[ \mu_{q(\beta)}^2 + \sigma_{q(\beta)}^2 \right]
+ \tfrac{1+d}{2} 
+ \tfrac{1}{2}\log|\bSigma_{q(\beta,\bu)}| \\
    & \ds \quad 
+ \tfrac{d}{2} 
+ \tfrac{1}{2}\log|\bSigma_{q(\bmu)}|
- \tfrac{nd}{2}\log(2\pi)
- \tfrac{d}{2}\log(\sigma_\mu^2)
- \tfrac{1}{2\sigma_\mu^{2}}\left[ \|\bmu_{q(\bmu)}\|^2 + \tr(\bSigma_{q(\bmu)}) \right]
 \\
    & \ds \quad 
+ \tfrac{\nu}{2}\log|\mPsi|
- \log\Gamma_d(\nu/2) 
- \tfrac{\nu+n}{2}\log|\mPsi_{q(\mSigma)}|
+ \tfrac{dn}{2}\log(2)
+ \log\Gamma_d((\nu + n)/2) \\
    &  \quad 
+ \sum_{i\in\sM} \tfrac{|\sM_i|}{2} 
+ \tfrac{|\sM_i|}{2}\log(2\pi) 
+ \frac{1}{2}\log|\bSigma_{q(\bd_{i,\sM_i})}|,
\end{array}
$$

\noindent with $\Gamma_p(\cdot)$ being the multivariate gamma function. 
Classification of $\widetilde{\bc}_i$ is finally performed through the decision rule $\text{sign}(\widetilde{\bc}_i^T\bmu^*_{q(\beta,\bu)})$.

\begin{algorithm} 
\noindent \caption{\textit{Iterative scheme for obtaining the parameters in the optimal densities for the variational Bayesian support vector machine with missing predictor values.}}
\label{alg5}
\begin{algorithmic}[1]
\REQUIRE $ \bmu_{q(\bC)}, \bmu_{q(\ba^{-1})}, \mu_{q({\sigma_{u}^{-2}})}, \bmu_{q(\bmu)}, \bSigma_{q(\bmu)}, \bmu_{q(\bSigma^{-1})}, \bSigma_{q(\bc_i)} (1 \leq i \leq n)$
\WHILE{the increase in $\log\underline{p}(\by,\bD_{\mbox{\tiny obs}};q)$ is significant}
    \STATE $\bW \leftarrow \diag(\bmu_{q(\ba^{-1})}) \quad ; \quad 
\widetilde{\bC} \leftarrow \bmu_{q(\bC)}$
    \STATE $\bSigma_{q(\beta,\vu)} \leftarrow 
\left[ 
\widetilde{\bC}^T\bW\widetilde{\bC} 
+ \{ \sum_{i=1}^n\mu_{q(a_i^{-1})} \bSigma_{q(\bc_i)} \}
+ \mbox{blockdiag}(\sigma_\beta^{-2},\mu_{q({\sigma_u^{-2}})} \bI_d)
\right]^{-1} 
\quad ; \quad 
\bmu_{q(\beta,\vu)} \leftarrow 
\bSigma_{q(\beta,\vu)}\widetilde{\bC}(\bI_n + \bW)\by$
    \STATE $\bOmega_{q(\beta,\bu)} \leftarrow \bSigma_{q(\beta,\bu)} + 
\bmu_{q(\beta,\bu)}\bmu_{q(\beta,\bu)}^T$
    \FOR{$i=1,\ldots,n$}
         \STATE 
$\bSigma_{q(\vd_{i,\mathcal{M}_i})} \leftarrow \left[ \bP_i^T\{
\bmu_{q(\bSigma^{-1})} + \mu_{q(a_i^{-1})}\bOmega_{q(\bu)}
\} \bP_i \right]^{-1}$
         \STATE
$
\begin{array}{rl}
\bmu_{q(\vd_{i,\mathcal{M}_i})} 
& \leftarrow
\bSigma_{q(\vd_{i,\mathcal{M}_i})} \bP_i^T\Big[
\bmu_{q(\bSigma^{-1})}\bmu_{q(\bmu)} 
+ y_i(1+\mu_{q(a_i^{-1})})\bmu_{q(\bu)}   
- \mu_{q(a_i^{-1})}[\bOmega_{q(\beta,\bu)}]_{-1,1} 
\\
& \quad  
- (\bmu_{q(\bSigma^{-1})} + \mu_{q(a_i^{-1})}\bOmega_{q(\bu)})\bQ_i\bQ_i^T\vd_i
\Big]\end{array}$ 
    \STATE $\bSigma_{q(\vd_i)} \leftarrow 
\bP_i \bSigma_{q(\vd_{i,\mathcal{M}_i})} \bP_i^T
\quad ; \quad 
\bSigma_{q(\bc_i)} \leftarrow \begin{bmatrix}
0        & \bzero_d^T \\
\bzero_d & \bSigma_{q(\vd_i)} \\
\end{bmatrix}$
    \STATE $\bmu_{q(\vd_i)} \leftarrow
\bP_i\bmu_{q(\vd_{i,\mathcal{M}_i})} + \bQ_i\bQ_i^T\vd_i
\quad ; \quad 
\bmu_{q(\bc_i)} \leftarrow [1,\bmu^T_{q(\vd_i)}]^T$
    \STATE 
$\chi_{q(a_i)} \leftarrow
(1-y_{i} \bmu_{q(\bc_i)}^T\bmu_{q(\beta,\bu)})^2 
+ \bmu_{q(\bc_i)}^T \bSigma_{q(\beta,\bu)} \bmu_{q(\bc_i)} 
+ \bmu_{q(\bu)}^T\bSigma_{q(\vd_i)}\bmu_{q(\bu)}
+ \mbox{tr}(\bSigma_{q(\bu)} \bSigma_{q(\vd_i)})$
\ENDFOR      
    \STATE $\bmu_{q(\ba^{-1})} \leftarrow \bchi_{q({\ba})}^{-1/2}$
    \STATE $\bSigma_{q(\bmu)} \leftarrow \left\{ 
\sigma_\mu^{-2}\bI_d +n\bmu_{q(\bSigma^{-1})} \right\}^{-1} 
\quad ; \quad 
\bmu_{q(\bmu)} \leftarrow \bSigma_{q(\bmu)}\bmu_{q(\bSigma^{-1})}
\left\{ \sum_{i=1}^{n}\bmu_{q(\vd_i)} \right\}$
    \STATE $\bPsi_{q(\bSigma)} \leftarrow \bPsi + n\bSigma_{q(\bmu)} 
+ \left( \sum_{i=1}^n 
(\bmu_{q(\vd_i)} - \bmu_{q(\bmu)})(\bmu_{q(\vd_i)} - \bmu_{q(\bmu)})^T
     + \bSigma_{q(\vd_i)}  \right) 
\quad ; \quad 
\bmu_{q(\bSigma^{-1})} \leftarrow (\nu + n) \bPsi_{q(\bSigma)}^{-1}$
    \STATE $B_{q(\sigma_u^2)} 
\leftarrow B_u+\tfrac{1}{2} \left[ \|\bmu_{q(\bu)} \|^2 + \text{tr}(\bSigma_{q(\bu)}) \right] \quad ; \quad 
\mu_{q({\sigma_u^{-2}})} \leftarrow (A_u+d/2)/B_{q(\sigma_u^2)}$
\ENDWHILE
\end{algorithmic}
\end{algorithm}

\section{Numerical experiments}
\label{sec:Numerical}

In this section, we present results for the traditional SVM approach, our VBSVM
approaches and MCMC based inference. 
For each of the VBSVM methods we have terminated the algorithm when the lower 
bound increases less than $10^{-10}$ between iterations. Unless otherwise 
specifically stated for each MCMC method 5000 burn-in samples are drawn followed 
by a further 5000 samples which are used for inference (no tinning is used).
For each model classification is performed using the posterior mean of the 
coefficient vector.

For non-simulated datasets we follow \cite{Kim:2009} for the assessment of 
classification performance. \cite{Kim:2009} recommends the repeated hold-out method 
because it has a reasonable computational cost against error variance trade-off. 
For this approach the data are split into 100 random training/test sets where the 
SVM is fit using 3/4 of the data and the classification error is calculated on the
remaining 1/4 of the data. The classification performance is then determined to 
be the average test balanced error rate (BER) over these 100 sets, where the BER is the average
of the error rates for both classes. The experiments are performed using an Intel Core i7-2760QM @ 2.40 GHz processor with 8 GBytes of RAM.

\subsection{Default SVM method}\label{res:default}

We will now describe our default method for selecting $\alpha$ in our SVM
formulation (\ref{eq:svmproblem}). We fit the traditional linear SVM using the 
\textsf{R} interface \texttt{e1071}, version 1.6 (Meyer, 2011) to the popular \texttt{LIBSVM} software.
To tune $\alpha$ we first select a grid of $\alpha$ values. For each 
particular $\alpha$ value we calculate the classification error based on 100
hold-out datasets by again splitting the training data. A second grid is 
then constructed centered around the $\alpha$ value with the smallest average 
test error and the process is repeated. The $\alpha$ value with the smallest 
test error from the second grid is selected for final testing on the original hold-out test set as described in Section \ref{sec:Numerical}.

\subsection{Penalty parameter inference}\label{res:1}

The first example is based on simulated data sets and compares the default SVM method, the VBSVM
approach described in Section \ref{sec:VB2} and the MCMC alternative (see 
Appendix A). The training data are generated according to
$$
\begin{array}{c}
\beta \sim N(0,1), 
\quad \bu \sim N(\bzero_d,\bI_d),
\quad \vd_i \sim N(\bzero_d,\bI_d), 
\quad q_i \sim \mbox{Bernoulli}(\mbox{expit}(\beta + \vd_i^T\bu)), 
\ 1 \leq i \leq n,
\end{array}
$$

\noindent with the final class labels calculated via $y_i = 2 q_i - 1$, $1\le 
i\le n$. We vary $n$ and $d$ over the sets $n \in \{ 100, 200, 500\}$ and 
$d\in\{10,50,100\}$. For each combination 200 random training data sets are
generated. 

\begin{figure}[!ht]
\begin{center}
\includegraphics[width=12cm,angle=-90]{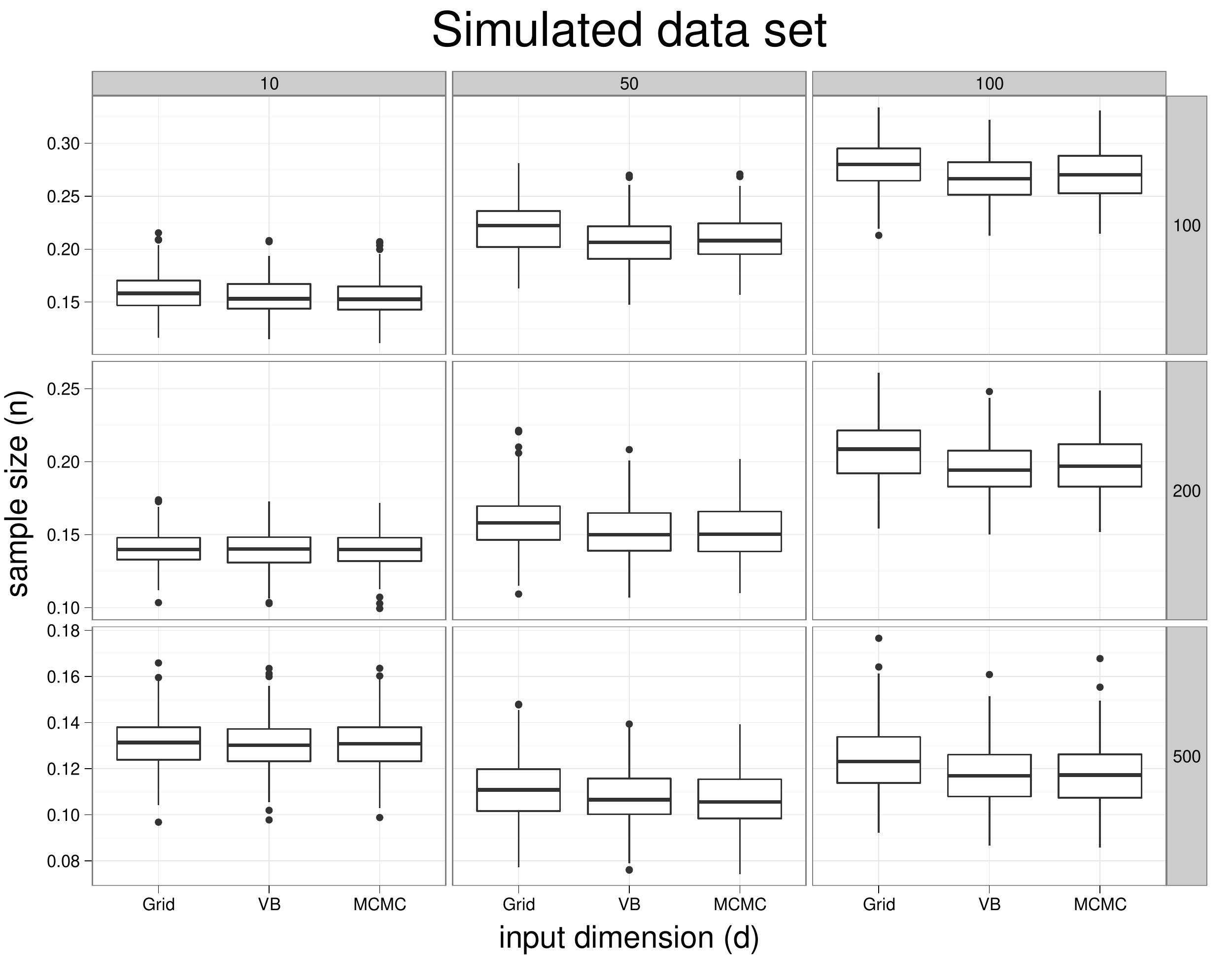}
\end{center}
\caption{Results of the simulated data described in Section \ref{res:1}. 
Balanced error rate for the default SVM approach (see Section 
\ref{res:default}), our VB method (see Section \ref{sec:VB2}) and MCMC inference 
for different values of input 
vector dimension $d\in\{10,50,100\}$ and training sample size $n\in\{100,200,500\}$.}
\label{classicalDataSimulation}
\end{figure}

Since this is a simulated dataset we can generate new test data to assess 
the performance for each method. For each of the 200 random training sets a
new independent test data set of 1000 input vectors is generated and the
BER is calculated. The 200 BERs on 
these independent test data are presented as boxplots in Figure 
\ref{classicalDataSimulation}. 

Figure \ref{classicalDataSimulation} shows that the performances of the VB 
algorithm and the MCMC approach are in general comparable. The use of the 
default 
SVM method achieves a similar classification performance compared to VB and MCMC 
for $d=10$. Increasing the training sample size tends to increase the 
performance of the grid approach, but its classification performance with 
respect to VB and MCMC is still slightly lower for $d=50$ and $d=100$. 
However, the big trade-off is in terms of computational efficiency. For example,
the default SVM method took on average 571.75 seconds for the case where $n = 200$ and $d = 10$ while the VB and MCMC methods took 1.68 seconds and 82.31 seconds respectively. Thus, 
while classification performances are similar our VBSVM approach is by far the 
fastest method and hence the method of choice here.

\subsection{Random intercept model}

We now show the effectiveness of our methodology for group correlated data. 
For this example we consider the toenail dataset of \cite{DEBACKER:1998}.
\cite{DEBACKER:1998} describe a clinical trial comparing the effectiveness of 
two oral antifungal treatments for toenail infection. Patients were randomly 
assigned to one of two treatment groups, one group receiving 250 mg per day of 
Terbinafine and the other group 200 mg per day of Itraconazole. Patients were 
evaluated at seven visits (approximately on weeks 0, 4, 8, 12, 24, 36, and 48) 
by recording the degree of onycholysis. In total, data from $m=294$ patients were 
available, comprising 1908 measurements. Only a dichotomized version of the 
longitudinally observed degree of onycholysis was included: 1500 observations of 
`absent' or `mild degree' belonging to the first group (with $y_{i,j} = -1$) and 408 
observations of a `moderate or severe degree'  of onycholysis belonging to the 
second group (with $y_{i,j} = +1$). 

Consider the classification problem where we wish to predict to which of these two 
groups the $j$th measurement from the $i$th patient belongs. Predictor variables
for the $(i,j)$th observation include visit time (${\tt visit}_{i,j}$), and 
treatment type (${\tt treat}_i$). We would expect the $y_{i,j}$ values to be
correlated within patients. Within a mixed model framework this correlation can 
be taken into account using a random intercept model. Hence, we consider the 
random intercept model as described in Section \ref{sec:VB2} with
$$
\bX = \left[ \begin{array}{cccc}
1 & {\tt visit}_{1,1} & {\tt treat}_1 & {\tt visit}_{1,1}\times{\tt treat}_1 \\
\vdots & \vdots & \vdots & \vdots  \\
1 & {\tt visit}_{1,n_1} & {\tt treat}_1 & {\tt visit}_{1,n_1}\times{\tt treat}_1 \\
1 & {\tt visit}_{2,1} & {\tt treat}_2 & {\tt visit}_{2,1}\times{\tt treat}_2 \\
\vdots & \vdots & \vdots & \vdots  \\
1 & {\tt visit}_{m,n_m} & {\tt treat}_m & {\tt visit}_{m,n_m}\times{\tt treat}_m \\
\end{array} \right] \quad \mbox{and} 
\quad \bZ = \left[ \begin{array}{cccc}
\bone_{n_1}  & \bzero_{n_1} & \cdots & \bzero_{n_1} \\
\bzero_{n_2} & \bone_{n_2}  & \cdots & \bzero_{n_2} \\
\vdots       & \vdots       & \ddots & \vdots \\
\bzero_{n_m} & \bzero_{n_m} & \cdots & \bone_{n_m} \\
\end{array} \right].
$$

\noindent Note that predictors have been standardized.

\begin{figure}[!ht]
\begin{center}
\includegraphics[width=10cm,angle=-90]{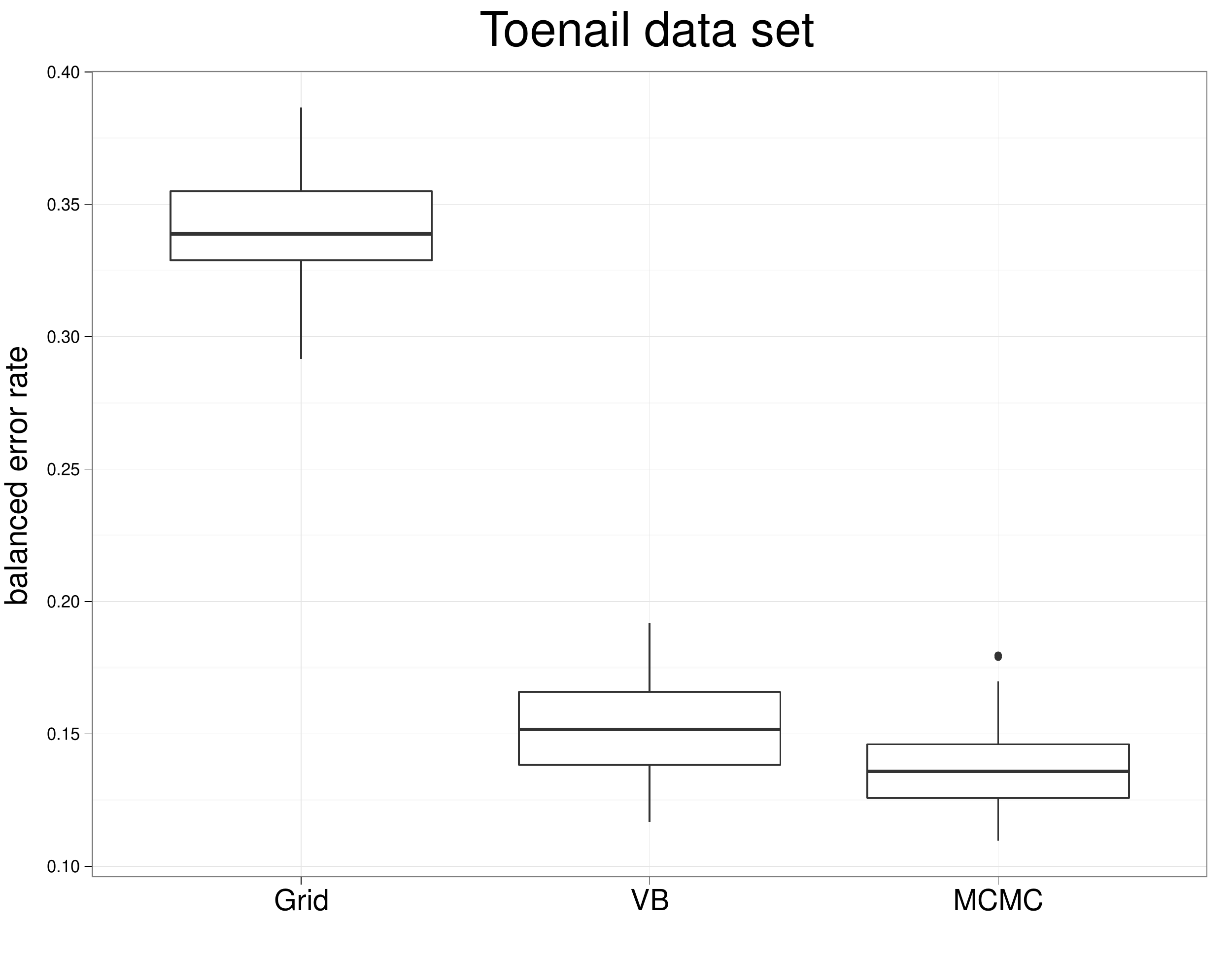}
\end{center}
\caption{\it Toenail data set. Balanced error rate for grid search, random intercept VB and random intercept MCMC inference.}
\label{toenail}
\end{figure}

The BERs for the 100 test sets are presented as boxplots in Figure \ref{toenail} 
and clearly illustrate the power of being able to incorporate such an effect in 
the model formulation of SVMs. The VB and MCMC methods with random intercepts 
show a better classification performance than traditional SVMs. The MCMC method 
tends to result in slightly lower BERs when compared to VB. In addition to the
classification performance increase there is also an efficiency increase for VB. 
The default SVM approach took on average 3668.70 seconds, VB took on average 
171.85 seconds and MCMC took on average 3597.12 seconds.

\subsection{Variable selection}

The example illustrates the use of a sparse prior for VB and MCMC inference on 
the spam data set \citep{FRANK:2010}. The spam data set was collected at the 
Hewlett-Packard Labs and consists of information from 4601 e-mails. A prediction 
vector of 57 variables was created for each e-mail and the goal is to predict 
whether the e-mail is spam or non-spam. The 57 variables include 54 percentages 
of word or character frequency in the e-mails and the 3 remaining predictors are 
related to the use of capital letters: the average length of uninterrupted 
sequences of capital letters, the length of the longest uninterrupted sequence 
of capital letters and the total number of capital letters in the e-mail. Note
that all predictors are standardized prior to analysis.

The model we consider is similar to that used by \cite{POLSON:2011} on the same
data set with $\rho$ also fixed at 0.01. Appendix B summarizes the full 
conditionals for our MCMC scheme and 
we use a burn-in size of 50000 samples and a retained set of 50000 samples 
because of slower mixing.

\begin{figure}[!ht]
\begin{center}
\subfigure[]{
\includegraphics[trim=3.2cm 2.5cm 3cm 4.5cm, clip=true,width=0.48\textwidth]{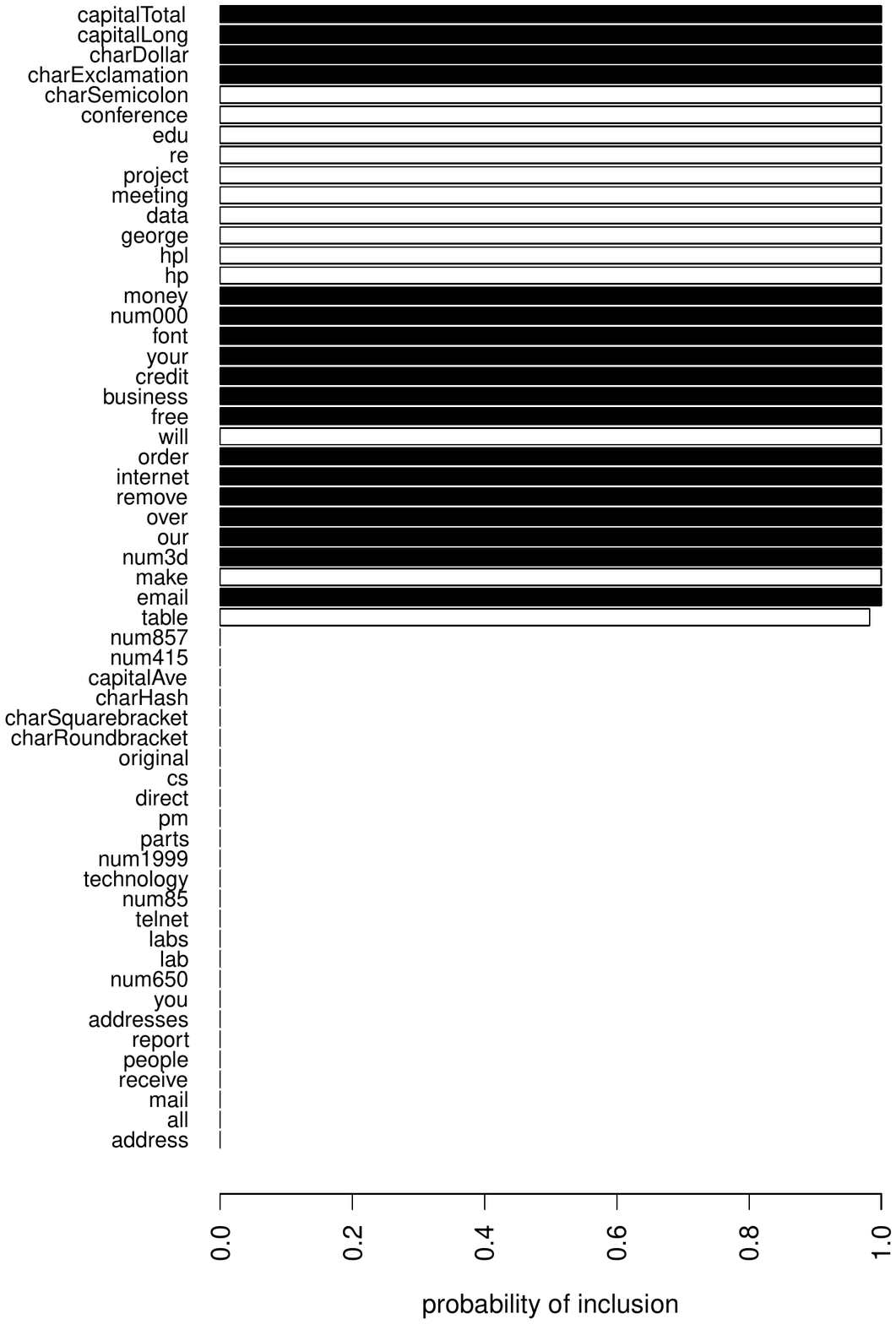}
}
\subfigure[]{
\includegraphics[trim=3.2cm 2.5cm 3cm 4.5cm, clip=true,width=0.48\textwidth]{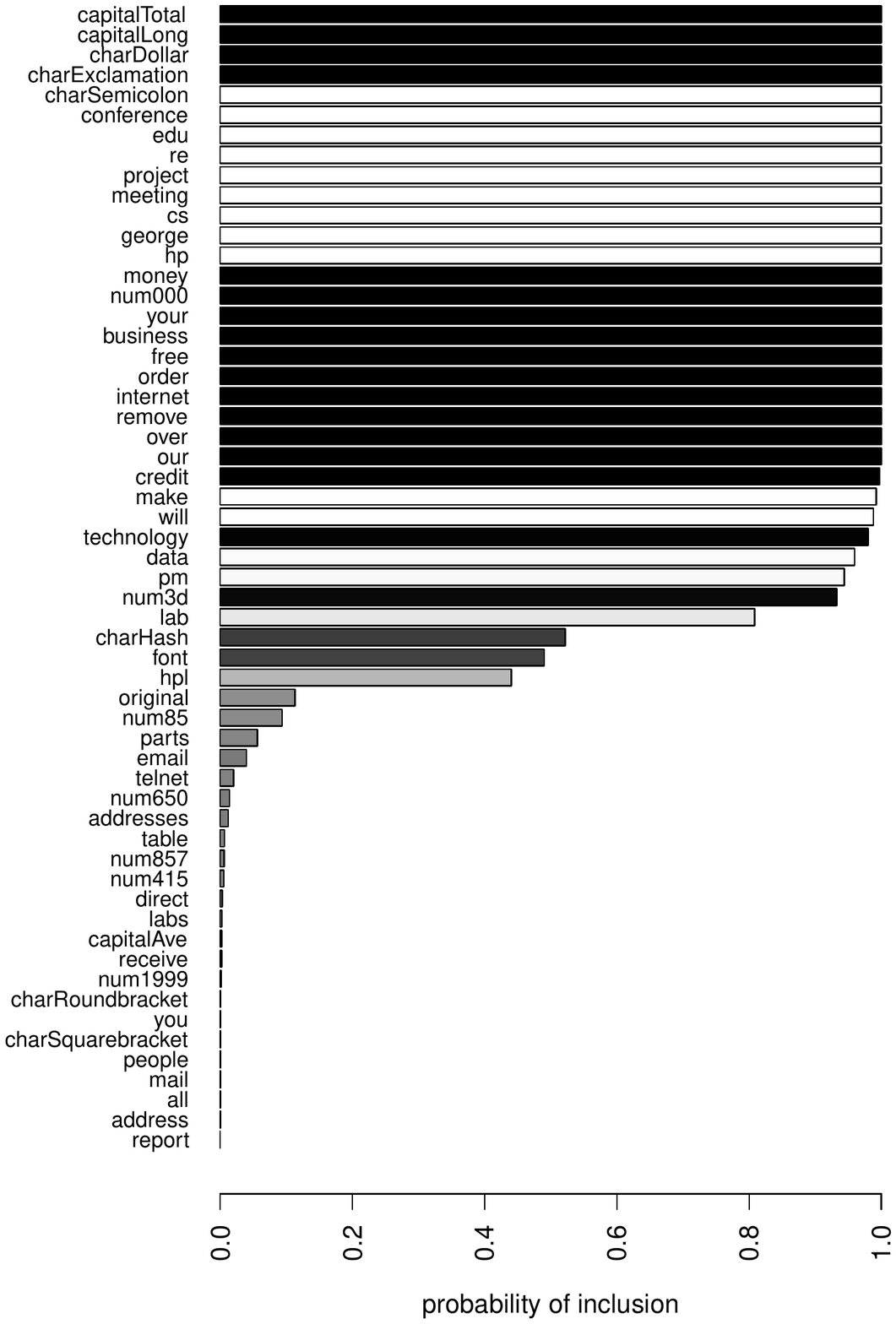}
}
\end{center}
\caption{\it Spam data set. Inclusion probabilities for VB and MCMC inference with a Laplace-zero prior and 
$\rho=0.01$ in (a) and (b), respectively. The bars are shaded in proportion to 
$P(v_k>0|\by)$, where darker means a greater probability for 
positive association with spam.}
\label{spam}
\end{figure}

Figure \ref{spam} illustrates the inclusion probabilities for each variable. 
Computing $P(v_k>0|\by)$ enables to visualize the results as black bars 
for variables that are strongly associated with the presence of spam, while the 
opposite is true for white bars. Although VB generates more extreme inclusion 
probabilities there exists good agreement between the VB and MCMC results. The 
24 variables that are almost certainly selected by MCMC match with the selected 
ones for VB, except for the variable {\tt cs}. Variables that are selected by VB 
correspond to MCMC selected ones, although {\tt hpl} and {\tt font} have slightly lower 
probabilities for MCMC. The two VB selected variables with smallest inclusion probabilities, i.e., {\tt email} and {\tt table} also have lower MCMC inclusion probabilities. In terms of speed VB is favorable over MCMC taking 76 minutes
compared to over 10 hours for MCMC.

%

\subsection{Missing predictor values}

The final example represents a classification problem where the interest lies in 
predicting the presence of significant coronary disease, which is defined as 
75\% or more diameter narrowing in at least one important coronary artery. The 
data consist of measurements from 3504 patients who were referred to Duke 
University Medical Center for chest pain, and are available through the Duke 
University Cardiovascular Disease Databank \citep{HARRELL:2001}. For each 
patient we have the {\tt age}, {\tt sex}, {\tt duration} of symptoms of coronary 
artery disease and the {\tt cholesterol} level as predictors. The latter two 
variables are log-transformed and the variables are standardized prior to 
analysis so that all variables are approximately standard normal. Importantly, 
the variable {\tt cholesterol} level has 1246 missing values.

The data set is repeatedly randomly split into a training and test part in such a way that all test cases have observed values for {\tt cholesterol}
level. For each split the BER on test set is computed for a traditional linear 
SVM and the missing predictor value VB (Algorithm \ref{alg5}) and MCMC (see Appendix C) approaches. Chapter 10 of \cite{HARRELL:2001} presents a logistic regression 
analysis of this data set where only the complete cases were retained from the 
original set. 

\begin{figure}[!ht]
\begin{center}
\includegraphics[width=10cm,angle=-90]{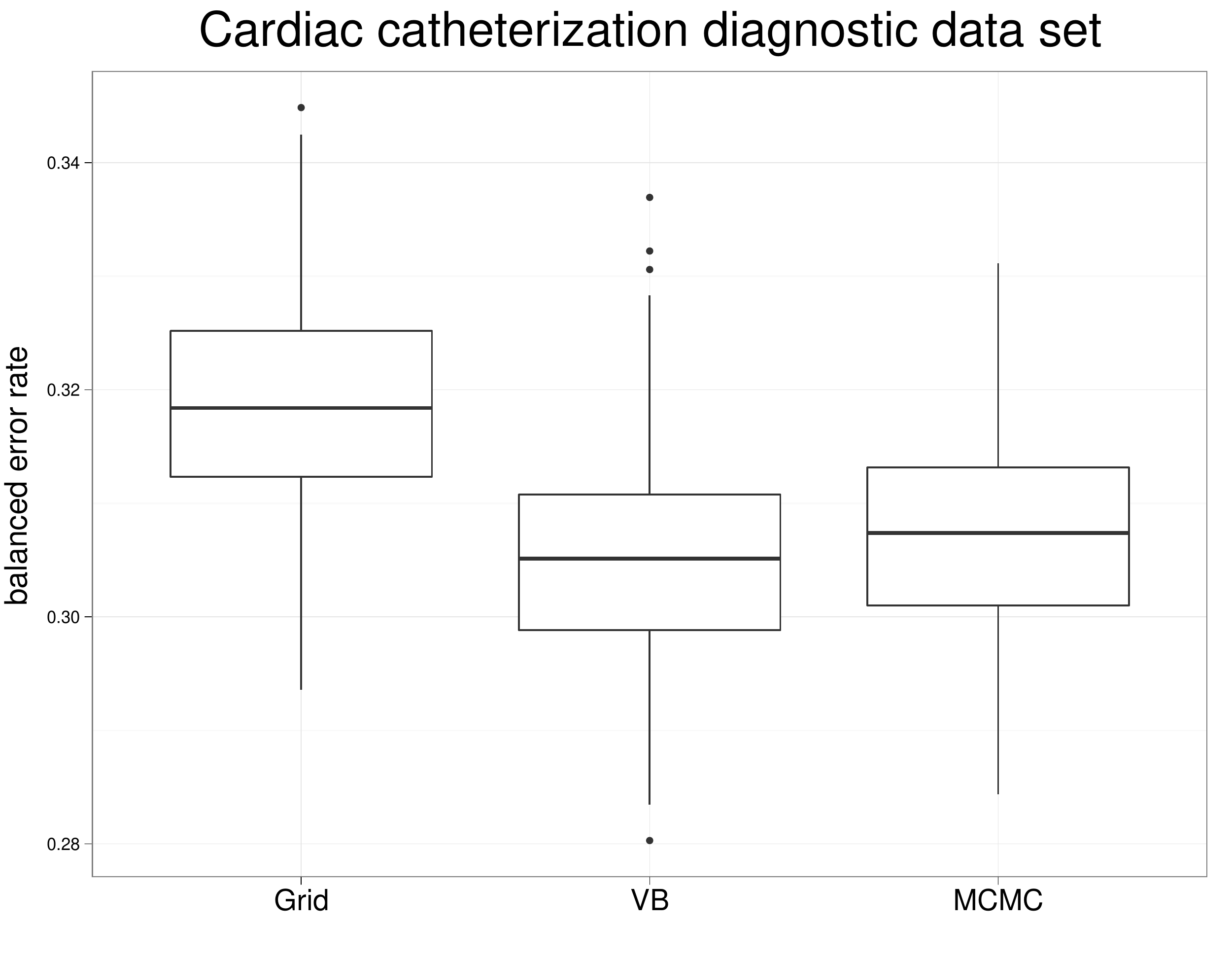}
\end{center}
\caption{\it Cardiac catheterization diagnostic data set. Balanced error rate for grid search with complete training cases, VB and MCMC inference in the presence of missing predictor values.}
\label{cathetrization}
\end{figure}

We compare our approaches against the default SVM approach which uses only complete 
cases for training. On the other hand, complete training cases and training cases with 
missing values for {\tt cholesterol} level are used for the VB approach and the MCMC scheme. Figure \ref{cathetrization} 
visualizes the boxplots of the BERs on test data for each of the three 
approaches. 

These results illustrate that methodology which allows to include input vectors
with missing values for training can yield better classification performance. 
In addition, the VB performance seems to be slightly better than the MCMC 
performance. Finally, the default SVM approach took on average 3109.40 seconds, the VB
approach took 1028.18 seconds and the MCMC approach took 1718.98. Hence, even
when missing data are present the VB approach is competitive both in terms of
classification performance and computational efficiency.

\section{Discussion}

We have developed a VB approach to SVM classification. We have 
shown that the approach is a unified framework for dealing with a variety of
complications typically difficult to deal with within a standard SVM framework.
For the examples that we present here our VBSVM methods have as good or better
classification performance than the standard SVM approach whilst remaining
computationally efficient.

\section*{Acknowledgments}

This research was partially supported by Australian Research Council Discovery Project DP110100061. 

\bibliographystyle{elsarticle-harv}
\bibliography{report}

\section*{Appendix A -- MCMC Scheme for Section 3.1}

\noindent The full conditionals for MCMC inference are
$$
\begin{array}{rl}
\bbeta,\bu|\mbox{rest}
    & \ds \sim 
N\Bigg\{ \left( 
\bC^T\mbox{diag}(\ba^{-1})\bC + 
\mbox{blockdiag}(\sigma_\beta^{-2}\bI_p,\sigma_u^{-2}\bI_m) 
\right)^{-1}\bC^T\bY(\bone_n + \ba^{-1}), \\
    & \ds \qquad \qquad \left( \bC^T\diag(\ba^{-1})\bC + 
\mbox{blockdiag}(\sigma_\beta^{-2}\mI_p,\sigma_u^{-2}\bI_m) \right)^{-1} 
\Bigg\}, \\
\sigma_u^2|\mbox{rest}
    & \sim \mbox{IG}\left(A_u + \tfrac{m}{2}, B_u + \tfrac{1}{2}\|\bu\|^2 
\right), \\
a_i|\mbox{rest}
    & \sim \mbox{GIG}\left(\tfrac{1}{2},1,(1-y_i(\bx_i^T\bbeta + \bz^T_i\bu))^2\right),
\end{array}
$$

\noindent where $y_i$ is the $i$th element of $\vy$ and $\vx_i$ and $\vz_i$
are the $i$th row of the matrices $\mX$ and $\mZ$ respectively. These can be 
used to implement a Gibbs sampling MCMC method.

\section*{Appendix B -- MCMC Scheme for Section 3.2}

\noindent The full conditionals for MCMC inference are
$$
\begin{array}{rl}
\bbeta,\bv|\mbox{rest}
    & \sim N\Big\{ \left( \widetilde{\bGamma}\bC^T\mbox{diag}(\ba^{-1})\bC\widetilde{\bGamma}  + \mbox{blockdiag}(\sigma_\beta^{-2}\mI_p,\sigma_u^{-2}\mbox{diag}(\bb)) \right)^{-1} \widetilde{\bGamma}\bC^T\bY(\bone_n + \ba^{-1}), \\
    & \ds \qquad \left( \widetilde{\bGamma}\bC^T\mbox{diag}(\ba^{-1})\bC\widetilde{\bGamma}  + \mbox{blockdiag}(\sigma_\beta^{-2}\mI_p,\sigma_u^{-2}\mbox{diag}(\bb)) \right)^{-1}\Big\}, \\
\sigma_u^2|\mbox{rest}
    & \sim \mbox{IG}\left( A_u + \tfrac{m}{2}, B_u + \tfrac{1}{2}\bv^T\mbox{diag}(\bb)\bv \right), \\
a_i|\mbox{rest}
    & \sim \mbox{GIG}(\tfrac{1}{2},1, 
(1-y_i(\bx_i^T\bbeta + \bz_i^T\bGamma\bv))^2), \\    
b_k|\mbox{rest}
    & \sim \mbox{Inverse-Gaussian}(\sigma_u/|v_k|,1), \\
\gamma_k|\mbox{rest}
    & \sim \mbox{Bernoulli}\Big[ \mbox{expit}\Big\{ 
\mbox{logit}(\rho) 
- \tfrac{1}{2}\bZ^T_k\mbox{diag}(\ba^{-1})\bZ_k v_k^2 
+ v_k\bZ^T_k\mbox{diag}(\bone_n + \ba^{-1}) \by \\
    & \qquad \qquad - v_k\bZ^T_k\mbox{diag}(\ba^{-1})(\bX\bbeta + \bZ_{-k} \mbox{diag}(\bgamma_{-k})\bv_{-k})
\Big\} \Big],  
\end{array}
$$

\noindent where $\widetilde{\bgamma}=[\bone^T_p,\bgamma^T]^T$ and $\widetilde{\bGamma} = \mbox{diag}(\widetilde{\bgamma})$.

\section*{Appendix C -- MCMC Scheme for Section 3.3}

\noindent The full conditionals for MCMC inference are

$$
\begin{array}{rl}
\beta,\bu|\mbox{rest}
    & \sim N\Big\{ \left( \bC^T\mbox{diag}(\ba^{-1})\bC + 
\mbox{blockdiag}(\sigma_\beta^{-2},\sigma_u^{-2}\bI_d) \right)^{-1}
\bC^T\bY(\bone_n + \ba^{-1}), \\
    & \ds \qquad \qquad 
\left( \bC^T\mbox{diag}(\ba^{-1})\bC + 
\mbox{blockdiag}(\sigma_\beta^{-2},\sigma_u^{-2}\bI_p) \right)^{-1} \Big\}, 
\\
\sigma_u^2|\mbox{rest}
    & \sim \mbox{IG}\left(A_u+\tfrac{d}{2},B_u+\tfrac{1}{2}\|\bu\|^2\right), 
\\
a_i|\mbox{rest} 
    & \sim \mbox{GIG}\left(\tfrac{1}{2},1,(1-y_i(\beta + \vd_i^T\bu))^2 \right),
\\
\bmu|\mbox{rest} 
    & \sim N\left[ \left\{
n\bSigma^{-1}+\sigma_\mu^{-2}\bI_d \right\}^{-1}n\bSigma^{-1}\overline{\vd},  
\left\{ n\bSigma^{-1} + \sigma_\mu^{-2}\bI_p \right\}^{-1}  \right], 
\\
\bSigma|\mbox{rest} 
    & \sim \mbox{IW}\left( \bPsi + \bD^T\bD - 2 n\overline{\vd}\bmu^T + n\bmu\bmu^T, \nu + n \right), 
\\
\vd_{i,\mathcal{M}_i}|\mbox{rest}
    & \sim N\Big[ \left\{
\bP_i^T\left[\bSigma^{-1} + a_i^{-1}\bu\bu^T\right]\bP_i \right\}^{-1}
 
\bP_i^T\Big[ \bSigma^{-1}\bmu + y_i(1 + a_i^{-1})\bu - a_i^{-1}\beta\bu
\\ 
    & \qquad \qquad  - \left[\bSigma^{-1} + a_i^{-1}\bu\bu^T\right]\bQ_i\bQ_i^T \vd_i
\Big],  

\left\{ \bP_i^T\left[\bSigma^{-1} + a_i^{-1}\bu\bu^T\right]\bP_i \right\}^{-1}
\Big],
\end{array}
$$
with $\overline{\vd} = \frac{1}{n} \sum_{i=1}^{n} \vd_i.$

\end{document}